\documentstyle[preprint,aps]{revtex}
\begin{document}
\draft

\title{Spin-dependent tunneling in metal-insulator-narrow gap 
semiconductor 
structures in a magnetic field}

\author{G.M.Minkov,  O.E.Rut,  and A.V.Germanenko}
\address{Institute of Physics and Applied Mathematics, Ural
University, Ekaterinburg 620083, Russia}
\date{\today}
\maketitle
\begin{abstract}
We present  results of tunneling studies of $\text{p-Hg}_{1-
x}\text{Cd}_x\text{Te-oxide-Al}$ structures  with $0.165<\text{x}<0.2$ in 
a magnetic field up to 6 T. The tunneling conductivity oscillations resulting 
from the Landau quantization of the energy spectrum in the semiconductor 
volume are investigated.
Under an in-plane magnetic field the amplitudes of tunneling conductivity 
maxima connected with the tunneling into {\em a} and {\em b} spin 
sublevels are found to differ substantionally from one another, and the 
amplitude ratio varying from structure to structure. To understand the cause 
of this behavior, the tunneling conductivity for this magnetic field 
orientation is calculated taking into consideration the multi-band energy 
spectrum.  It is shown that the contributions of the different spin-sublevels to 
the tunneling conductivity are dissimilar and the relationship between them  
depends strongly on the value of surface potential.
\end{abstract}

\narrowtext
\section{Introduction}
\label{sec:intro}

	The electron tunneling in metal -- insulator -- semiconductor (MIS) 
structures  is a useful tool for studying the energy spectrum of two- and 
three-dimensional states, the interaction of  carriers with the 
semiconductor-insulator boundary, the interaction with elementary 
excitations such as 
optical phonons, magnons, etc.\cite{1}  The tunneling in MIS 
structures based on wide-gap semiconductors, for example GaAs has been 
studied in detail.\cite{2,2a,3,4,5,6,7,8,9,10} In these semiconductors the 
electron effective $g$-factor is 
small and spin-dependent effects in tunneling do not manifest 
themselves. The opposite situation is observed in  HgTe-type 
semiconductors with an inverted energy spectrum, where the electrons 
are the carriers of the fourfold degenerate (at $k=0$) band.  The difference 
of contributions by different electron spin states to various 
phenomena for such materials was discussed by a number of 
authors.\cite{11,12} It has been found experimentally that with the  
magnetic field parallel to the current, the amplitudes of Shubnikov-de-Haas 
oscillations connected with the different spin states differ largely.\cite{11} 
An analogous difference in tunneling conductivity in structures based on 
inverted-spectrum semiconductors has been observed 
with magnetic field perpendicular to the normal to the plane of a 
tunneling structure ($B\perp n$).\cite{12} 
With this magnetic field orientation, the 
energy of the states depends not only on the quantum number, $n$, and 
the quasimomentum in the direction of the magnetic field, $k_B$, but also 
oscillation center -- bounadry distance, $x_0$, and 
this $x_0$ dependence of the energy is essential for oscillation 
amplitudes of the tunneling conductivity in the magnetic field. 
It is well known that for a twofold degenerate parabolic band the energy of 
an electron increases monotonically with decreasing $x_0$. This $E$-$x_0$ 
relation is the same for both spin states. Consequently,  the 
contributions to the tunneling conductivity, that come from 
tunneling into different spin states, should be close in magnitude in MIS 
structures based on such materials. In terms of the one-band energy 
spectrum model, magnetotunneling under an in-plane magnetic field was 
considered in Refs. \onlinecite{7,8,9,10}. In semiconductors with a complex 
spectrum, the behavior of the different spin states near the boundary 
may  differ significantly and it should lead to a distinction in tunneling 
into different spin states.  Thus the behavior of  the Landau levels in 
the vicinity of the boundary in the semiconductors with the energy 
spectrum described by the Dirac Hamiltonian was calculated in Ref. 
\onlinecite{13}. 
It was shown that in a symmetric band-inverted junction the 
Landau eigenenergies vary nonmonotonically with the orbit center -- 
structure interface distance and this behavior various for different spin 
states. The same problem was considered also for semiconductors 
with the inverted spectrum, described by the Luttinger Hamiltonian.\cite{14} 
The behavior of the Landau levels near the boundary is different for 
different spin states in this case, too, and this fact was drawn on to 
explain the significant difference of the tunneling conductivity oscillation 
amplitudes associated with tunneling into different spin 
states in inverted semiconductor $\text{p-Hg}_{1-
x}\text{Cd}_x\text{Te}$-insulator-metal tunneling structures at $B\perp 
n$. The fact that the electrons in inverted semiconductors were the carriers 
of the fourfold degenerate $\Gamma_8$ band was essential to the 
explanation of the observed peculiarities. 

	Surprisingly, the  similar 
peculiarities of the tunneling conductivity into different spin states at 
$B\perp n$ are observed for some tunneling 
structures based on narrow-gap semiconductors, where the electrons 
are the carriers of the twofold degenerate $\Gamma_6$ band.  However, the 
ratio of the tunneling conductivity oscillation amplitudes associated with 
tunneling into different spin states varies widely for various tunnel 
structures. One 
possibility is that the surface potential at the  semiconductor-insulator 
boundary is different in these tunneling structures. Such peculiarities 
of the tunneling into different spin states are the subject of the present 
paper.

	The article is organized as follows. Section \ref{two} presents and 
analyzes an experimental data for tunneling conductivity oscillations with 
various magnetic field orientations. Section \ref{three}  is devoted to the 
calculation of the behavior of the Landau levels near the surface of a 
Kane spectrum semiconductor.  The basic equations and 
formalism are spelled out. Results of the calculation of tunneling 
conductivity oscillations and a comparison with  experimental data are 
presented in Section \ref{four}.	
	
\section{Experimental results}
\label{two}

	We have investigated the tunneling conductivity $\sigma_d=dj/dV$ 
and its derivative $d\sigma_d/dV$ as functions of bias $V$ and magnetic 
field in p-$\text{{Hg}}_{1-x}\text{Cd}_{x}\text{Te}$-oxide-Al 
($0.165<x<0.21$) structures with $10<E_g<100$ meV in  
magnetic fields up to 6 T at a temperature of 4.2 K.  Tunnel 
structures were fabricated on single-crystal   p-$\text{{Hg}}_{1-
x}\text{Cd}_{x}\text{Te}$  with 
uncompensated acceptor concentration  $N_A-N_D=(0.8-2)\times10^{18}$ 
$\text{cm}^{-3}$. 
The doping level was determined from an analysis of galvanomagnetic 
phenomena in the temperature range 4.2 - 80 K.  The procedure for 
the fabrication of tunneling structures was described in our previous 
papers.\cite{15,16} The resistance of our structures was  $0.01-1$ kOhm. 
Several tunnel structures prepared on each sample were 
investigated. 

	The energy diagram of the tunneling contact is shown in Fig. 
\ref{fig1}. 
The bias shifts the Fermi level of the metal relative to that of the 
semiconductor ($E_F$) and the tunneling conductivity is proportional to the 
density of states in the semiconductor at energy $E=E_F+eV$. The 
magnetic field quantizes the energy spectrum of the semiconductor 
and the tunneling conductivity becomes an oscillatory function of  
magnetic field and bias. Typical oscillation pictures of the 
tunneling conductivity in a magnetic field at a fixed bias  for two different  
structures  are shown in Fig.\ref{fig2}. It is seen that (i)
for both  structures the 
oscillation maxima are split,  and at $B \parallel n$ the amplitudes of the 
components are comparable in magnitude; (ii) 
the amplitudes of the maxima and magnetic field positions exhibit a rather 
complicated behavior in tilted magnetic 
field; (iii) at $B\perp n$ the ratio of the amplitudes of the components are 
close in magnitude for structure I and they differ significantly  for
structure II, so  that high-field components are hard to
resolve at this  magnetic field  orientation.  Similar peculiarities of 
oscillation maxima amplitudes are observed in the bias dependences of 
$dj/dV$ at a fixed magnetic field (Fig. \ref{fig3}). 

	A Fourier analysis of the oscillations shows that  the $\sigma_d$ vs 
$B^{-1}$  
curves have two fundamental fields, $B_1$ and $B_2=2B_1$.  The values 
of fundamental fields  do not vary with angle between $B$ and $n$,  only 
the amplitude of Fourier component $B_2$ decreases when $\Theta 
\rightarrow 90^{\circ}$ so 
that it practically disappears for 
structure II. Thus, this behavior of the oscillation provides evidence that 
the tunneling conductivity  oscillations for both structures are 
due to tunneling into bulk Landau levels, splitting of the oscillation 
maxima corresponds to the spin splitting of the Landau levels, and the
probability of tunneling into spin-up and spin-down states for 
structure II is remarkably different at $B \perp n$. 

	The fundamental field $B_1$ at bias $V$ is determined by the value 
of quasimomentum at the energy $E=E_F+eV$ , $B_{1}^{-1}=2e/c\hbar 
k^2$, so that measurements at various biases make it possible to obtain  the 
dispersion law $E(k)$ for a wide energy range. \cite{17,18} For the  
structures 
investigated, the experimental data are in good agreement with the 
$E(k)$ curve calculated in terms of the Kane model with parameters 
$P=8.2\times10^{-8}$ eV cm, $E_g=80$ meV, $\Delta=\infty$.

	It is well known that the phenomena occurring in a tilted magnetic 
field are a very complicated problem. We will therefore restrict our attention 
to the 
results at $B \parallel n$ and $B \perp n$, namely, the ratio between the 
oscillation amplitudes for tunneling into different spin states and 
the relationship of this ratio to the structure parameters. 

	Let us consider the tunneling current and tunneling 
conductivity at $B\parallel n \parallel Ox$ and $B\perp n\parallel Ox$.  
For  a metal-insulator-semiconductor 
structure, the Fermi energy of the metal is much greater than that of the 
semiconductor; therefore, upon summation over all metal states, the 
tunneling current for a structure with low barrier transparency can be 
written in the form
\begin{equation}
j(V)\propto D \sum_{s}\left(\hat{v}\Psi\right)^{2}_{x=0}
\label{eq1}
\end{equation}
Here $D$ is the barrier transparency,   $\hat{v}$, $\Psi$  are the electron 
velocity operator and the wave function in the semiconductor, and $x=0$ is the coordinate of the boundary. The summation runs over all states in the semiconductor, from energy  $E_F$  to $E_F+eV$. 

	In a magnetic field, the semiconductor's states  are defined by 
three quantum numbers: Landau level number $n$, oscillator center 
position, and the quasimomentum component in the direction of the 
magnetic field. At $B\parallel n \parallel Ox$  $(B=B_x)$ the quantity 
$(\hat{v}_x\Psi_n)^2$ depends on $k_x$ only, and going to integration 
over energy we obtain 
\begin{eqnarray}
j(V)&\propto&D\sum_{n}\int\left(\hat{v}\Psi_{n}\right)^2_{x=0}dk_x 
\nonumber \\
     &=&
D\sum_{n}\int_{E_F}^{E_F+eV}\left(\hat{v}\Psi_{n}\right)^2_{x=0}\frac{dk_
x}
{dE}dE
\label{eq2}
\end{eqnarray}	
\begin{equation}
\sigma_d(V)\equiv\frac{dj}{dV}\propto D \left(\hat{v}\Psi_{n}\right)^2_{x=0}
\frac{dk_x}{dE}|_{E=E_F+eV}
\label{eq3}
\end{equation}	

In  the magnetic field $B \perp n \parallel Ox$. the Landau eigenenergies 
are a function of  the orbit center-barrier distance $x_0$ and $k_B$, so the 
square of the velocity at the boundary is also a function of $x_0$ and 
$k_B$,
\begin{eqnarray}
j(V)&\propto&D\sum_{n}\int\int\left(\hat{v}\Psi_{n}\right)^2_{x=0}dx_0dk_B
\nonumber \\
&=&
D\sum_{n}\int\int\left(\hat{v}\Psi_{n}\right)^2_{x=0}dk_B\left(\frac{dE}
{dx_0}\right)^{-1}dE
\label{eq4}
\end{eqnarray}	
\begin{equation}
\sigma_d(V)\propto\int_{0}^{k_{B}^{max}}\left(\hat{v}\Psi_{n}\right)^2_{x=0}
\left(\frac{dE}{dx_0}\right)^{-1}dk_B|_{E=E_F+eV}
\label{eq5}
\end{equation}	

Thus, to calculate the tunneling conductivity at $B\perp n$, one needs to 
calculate the  distance dependencies of the velocity and energy 
near the boundary. This problem for a semiconductor with a Kane 
spectrum is considered in the next section.

\section {The Landau levels near the boundary at 
{\bf B}$\perp$\lowercase{\bf n}}

\label{three}
	To calculate the behavior of the Landau levels near the 
boundary at $B \perp n$ and the tunneling current for this magnetic field 
orientation, let us consider an insulator-semiconductor structure with an 
abrupt interface (Fig. \ref{fig4}).  We assume that the insulator has the same 
energy structure as the semiconductor with the same momentum matrix 
element $P$ but with an energy gap value much 
greater than that for a semiconductor. \cite{18a} The parameters $D_c$, and $D_v$ 
are  
the conduction and valence band offsets, respectively. For such a structure 
we can solve the Schr\"odinger equation for $x<0$ (i.e., for an insulator) 
and $x>0$ (i.e., for a semiconductor) independently. Matching these 
solutions at $x=0$ in the required way, we find the solution for entire 
structure. 
	
	We use the Kane energy spectrum model for a semiconductor and an 
insulator. The interaction with remote and spin-orbit split $\Gamma_7$ 
bands is neglected. We choose the direction $x$ to be to 
normal to the interface and $z$ to be aligned with the magnetic 
field($B=(0,0,B)$, $B\perp n$).  The electromagnetic vector potential $A$ is 
chosen 
as $A=(0,B(x-x_0),0)$. In this case the components of the momentum 
operator are $\hat{k}_x=-i\frac{\partial}{\partial x}$, $\hat{k}_y=-L^{-
2}(x-x_0)$,     $\hat{k}_z=k_z=k_B$. Here $L=\sqrt{\hbar c/eB}$ is 
the magnetic length. 
To start with we wish to consider this problem for $k_z=k_B=0$. In this 
case  the Schr\"odinger equation  $\hat{H}\Psi=E\Psi$  splits up into two 
independent equations corresponding to the Landau spin sublevels, 
designated as {\em a}-  and {\em b}-sets. For the {\em a}-set these 
equations read as 
\widetext
\begin{equation}
\hat{H}^a\Psi^a=E\Psi^a,
\label{eq6}
\end{equation}
where
\begin{displaymath}
\hat{H}^a=
\left(
\begin{array}{ccc}
    E_g & -\frac{i}{\sqrt{6}}\frac{P}{L}
	\left(\frac{\partial}{\partial\xi}+(\xi-\xi_0)\right)& 
	-\frac{i}{\sqrt{2}}\frac{P}{L}
	\left(\frac{\partial}{\partial\xi}-(\xi-\xi_0)\right) \\
    -\frac{i}{\sqrt{6}}\frac{P}{L}
	\left(\frac{\partial}{\partial\xi}-(\xi-\xi_0)\right) &
	0 &
	0 \\
    -\frac{i}{\sqrt{2}}\frac{P}{L}
	\left(\frac{\partial}{\partial\xi}+(\xi-\xi_0)\right)  &
	0 &
	0
\end{array}
\right),
\end{displaymath}
and for the {\em b}-set,
\begin{equation}
\hat{H}^b\Psi^b=E\Psi^b,
\label{eq7}
\end{equation}
where

\begin{displaymath}
\hat{H}^b=
\left(
\begin{array}{ccc}
    E_g & \frac{i}{\sqrt{2}}\frac{P}{L}
	\left(\frac{\partial}{\partial\xi}+(\xi-\xi_0)\right)& 
	\frac{i}{\sqrt{6}}\frac{P}{L}
	\left(\frac{\partial}{\partial\xi}-(\xi-\xi_0)\right) \\
    \frac{i}{\sqrt{2}}\frac{P}{L}
	\left(\frac{\partial}{\partial\xi}-(\xi-\xi_0)\right) &
	0 &
	0 \\
    \frac{i}{\sqrt{6}}\frac{P}{L}
	\left(\frac{\partial}{\partial\xi}+(\xi-\xi_0)\right)  &
	0 &
	0
\end{array}
\right).
\end{displaymath}
\narrowtext
Here $\xi=x/L$, and $\xi_0=x_0/L$ are dimensionless coordinates, 
energies $E$ and $E_g$ are measured from the top of the semiconductor 
valence band. We seek solutions of (\ref{eq6}) and (\ref{eq7}) in the form
\begin{equation}
\Psi^\pm=
\left(
\begin{array}{ccc}
c_1u^\pm\left(a_1,(\xi-\xi_0)^2\right)\\
c_2u^\pm\left(a_2,(\xi-\xi_0)^2\right)\\
c_3u^\pm\left(a_3,(\xi-\xi_0)^2\right)\\
\end{array}
\right)e^{-\frac{(\xi-\xi_0)^2}{2}}
\label{eq8}
\end{equation}
where 
\begin{eqnarray}
u^\pm(a,x^2)&=&\frac{\sqrt{\pi}}{\Gamma(p+\frac{1}{2})}M(a,\frac{1}{
2},x^2) \nonumber \\
& \pm &\frac{2\sqrt{\pi}}{\Gamma(p)}xM(a+\frac{1}{2},\frac{3}{2},x^2),
\nonumber
\end{eqnarray}
and $M(a,b,y)$ is a confluent hypergeometric function.\cite{19}  Using the 
properties of the confluent hypergeometric function,  one can write the  
particular solutions for the {\em a}-set
\begin{equation}
\Psi^{\pm}_{a}=
\left(
\begin{array}{ccc}
ELP^{-1}u^\pm\left(a,(\xi-\xi_0)^2\right)\\
i\sqrt\frac{2}{3}u^\pm\left(a-\frac{1}{2},(\xi-\xi_0)^2\right)\\
i\sqrt{2}au^\pm\left(a+\frac{1}{2},(\xi-\xi_0)^2\right)\\
\end{array}
\right)e^{-\frac{(\xi-\xi_0)^2}{2}},
\label{eq9}
\end{equation}
where 
\begin{displaymath}
a=\frac{3}{8}\left(\left(\frac{L}{P}\right)^{2}E(E_g-E)+\frac{1}{3}\right)
\end{displaymath}
and for the {\em b}-set,
\begin{eqnarray}
\Psi^{\pm}_{b}=&\nonumber \\
&\left(
\begin{array}{ccc}
ELP^{-1}u^\pm\left(a,(\xi-\xi_0)^2\right)\\
-i\sqrt{2}u^\pm\left(a-\frac{1}{2},(\xi-\xi_0)^2\right)\\
-i\sqrt{\frac{2}{3}}au^\pm\left(a+\frac{1}{2},(\xi-\xi_0)^2\right)\\
\end{array}
\right)e^{-\frac{(\xi-\xi_0)^2}{2}},
\label{eq10}
\end{eqnarray}
where 
\begin{displaymath}
a=\frac{3}{8}\left(\left(\frac{L}{P}\right)^{2}E(E_g-E)+1\right).
\end{displaymath}
The solutions for the insulator that correspond to the same energy as the 
solution for 
semiconductor are given by (\ref{eq9}), (\ref{eq10}), where $E$ and $E_g$ 
are changed to 
$E+D_v$ and $E_g+D_c+D_v$, respectively.

The general solution for $x>0$ $(R)$ and $x<0$ $(L)$ 
for the {\em a}-set is
\begin{eqnarray}
\Psi_{a}^{R,L}(x)=
\left[
\begin{array}{ccc}
\psi^{1}_{a}(x) \\
\psi^{2}_{a}(x) \\
\psi^{3}_{a}(x)
\end{array}
\right]_{R,L}&=&A^{R,L}\left[\Psi_{a}^{+}\right]_{R,L}\nonumber \\
&&+B^{R,L}\left[\Psi_{a}^{-}\right]_{R,L}
\label{eq11}
\end{eqnarray}

The general solution for the {\em b}-set is similar in form. 
The quantity $\Psi^R$  has to converge at  $x\rightarrow+\infty$ and 
$\Psi^L$  has to  converge at  $x\rightarrow-\infty$, so
we must set $A^R=B^L=0$.  Matching conditions are obtained 
by integrating the Schr\"odinger equation across the boundary. 
For the {\em a}-set these conditions are:
\begin{mathletters}
\label{eq12}
\begin{equation}
\left[\psi^{1}_{a}(0)\right]_{L}=\left[\psi^{1}_{a}(0)\right]_{R}
\end{equation}
\begin{equation}
\left[\psi^{2}_{a}(0)+\sqrt{3}\psi^{3}_{a}(0)\right]_{L}=
\left[\psi^{2}_{a}(0)+\sqrt{3}\psi^{3}_{a}(0)\right]_{R}
\end{equation}
\end{mathletters}
and for the {\em b}-set,
\begin{mathletters}
\label{eq13}
\begin{equation}
\left[\psi^{1}_{b}(0)\right]_{L}=\left[\psi^{1}_{b}(0)\right]_{R}
\end{equation}
\begin{equation}
\left[\sqrt{3}\psi^{2}_{b}(0)+\psi^{3}_{b}(0)\right]_{L}=
\left[\sqrt{3}\psi^{2}_{b}(0)+\psi^{3}_{b}(0)\right]_{R}
\end{equation}
\end{mathletters}
These matching conditions give the secular equation for calculating 
the $x_0$ dependence of energy. This problem was  solved 
numerically with parameters corresponding to the samples 
investigated ($E_g=80$ meV, $P=8.2\times10^{-8}$ eV cm) and $D_c=2$ 
eV, $D_v=1$ eV.\cite{15}  Such curves for the {\em a}- and {\em b}-sets 
are presented in Fig. \ref{fig5} in dimensionless coordinates ($x/L$,$E/\hbar\omega_n$), where $\omega_n=eB/m_n c$, and $m_n$ is the electron effective mass at the conduction band bottom. A pronounced difference between the spin 
sublevels is seen. This is the result of the spin-orbit interaction with the 
boundary.

\section{Discussion}
\label{four} 

It is evident from Eq.(\ref{eq5}) that to calculate the tunneling 
conductivity, we need the $x_0$ dependence of the velocity squared at 
the boundary.  With the definition of the velocity operator 
$\hat{v}=i/\hbar[Hx]$ we 
have calculated $(\hat{v}\Psi)^2$ for the levels presented in Fig.\ 
\ref{fig5}, and as an illustration these results for sublevels  $\text{\em 
a}_1$, and $\text{\em b}_1$ are plotted in Fig.\ \ref{fig6}. 
In calculating the tunneling conductivity, we have assumed  that 
the $x_0$ dependencies of energy and $(\hat{v}\Psi)^2$  have the same 
form at any 
$k_B$ (Fig. \ \ref{fig5}, \ref{fig6}). In Fig.\ \ref{fig7} we show the 
contributions that the tunneling 
into different Landau sublevels makes to the tunneling conductivity 
as calculated from Eq.(\ref{eq5}). Note that tunneling into {\em b}-levels is 
more effective in consequence of differences in the $E$ vs $x_0$ and 
$(\hat{v}\Psi)^2$ vs $x_0$ curves for  {\em a} and {\em b} sublevels. The 
model 
discussed here has only two parameters, $D_c$, and $D_v$, and the results 
have only a weak dependence on these parameters. Finally, summing 
the contributions to the tunneling conductivity by the different Landau 
sublevels and subtracting the monotone component, we obtain  the 
oscillatory part of the tunneling conductivity presented in Fig.\ \ref{fig8}. 
For comparison, we show in Fig.\ \ref{fig8} also the oscillations of the 
tunneling 
conductivity of the same structure at $B\parallel n$, which are calculated in 
the same way 
as in Ref. \onlinecite{16}. It is seen that at $B\parallel n$
the amplitudes of  the tunneling 
maxima connected with tunneling into {\em a} and {\em b} sublevels are 
close in magnitude.  

	Inspection of the experimental and calculated results (Figs.  
\ref{fig3}, \ref{fig8}) shows that the relationship among the oscillation 
amplitudes at $B\parallel n$ agree well with experimental data for both 
structures. Whereas at $B\perp n$ 
the ratio of the {\em a}-maximum amplitude to the {\em b}-maximum 
amplitude  is appreciably greater than the calculated value  for structure I, 
but much
less for structure II. One conceivable reason for this discrepancy is 
the presence of an attractive or a repulsive potential in the 
semiconductor near the semiconductor-insulator interface. In order to 
verify this assumption the above calculation was carried out  
for structures with a model potential which we suppose as a rectangular 
well (or a barrier) of width $d$ and value $U$.  In this case the wave 
functions in three regions, $x<0$, $0<x<d$ and $x>d$, have the same form 
(\ref{eq11}). It is essential that for $0<x<d$ both terms in Eq.(\ref{eq11}) 
should be taken into account. The requirement that the boundary conditions  
(\ref{eq12}) and (\ref{eq13}) be fulfilled at 
$x=0$ and  at $x=d$ gives a secular equation for calculating the $x_0$
dependence of energy. The results of calculations for the surface well 
($eU=-70$ meV, $d=25$ \AA) and the barrier ($eU=70$ meV, $d=25$ 
\AA) are 
presented in Figs. \ref{fig9}, and \ref{fig10}.  The parameters of the 
potential are so adjusted that the well cannot hold the 2D states. It is seen 
that the effect of the 
surface potential is different for the {\em a}- and {\em b}-sublevels. As a 
result, the 
ratio between amplitudes of tunneling conductivity maxima
associated with tunneling into  {\em a}- and {\em b}-sublevels is strongly 
dependent on the surface potential. For the repulsive potential the 
amplitudes of {\em a}- and {\em b}-maxima can be close in magnitude (Fig.\ 
\ref{fig10}c), as is experimentally observed for structure I (Fig.\ \ref{fig3}a), while
for a sufficiently  large attractive potential only  {\em
b}-maxima are visible (Fig.\  \ref{fig10}a), as is
experimentally observed for structure II (Fig.\  
\ref{fig3}b). Note that for  $B\parallel n$, the  calculated
and experimentally observed amplitudes  of the tunneling conductivity 
maxima associated with tunneling into  {\em a}- 
and {\em b}-sublevels are close in magnitude, for both the attractive and 
the repulsive potential. 

	Thus, the ratio of amplitudes of maxima due to tunneling into {\em 
a} and {\em b} sublevels at $B \perp n$ depends on the surface potential 
and provides 
a way for estimating the value of the surface potential under
conditions where the surface potential is small and cannot localize the 2D states.

	As mentioned above (ii), the amplitudes of the maxima and the 
magnetic field positions in a tilted magnetic 
field exhibit a rather complicated behavior (Fig.\ \ref{fig2}).
This is clearly  seen for maxima observed in a high magnetic
field. The positions of the  maxima vary nonmonotonically with angle and with
$\Theta \simeq 50^{\circ}$ their amplitudes decrease drastically. Basically, 
the same behavior has been observed for all the structures investigated. 

	In addition, we would like to point out  one further peculiarity 
of tunneling conductivity in a tilted magnetic field. It is most 
pronounced in bias dependencies of $d\sigma_d/dV$ in the structures 
based 
on narrower-gap $\text{Hg}_{1-x}\text{Cd}_x\text{Te}$ (Fig.\ 
\ref{fig11}). A fine structure of maxima associated with tunneling into {\em 
b}-sublevels arises with magnetic field orientations close to $B\perp n$ 
($85^\circ>\Theta>50^\circ$). All {\em b}-maxima have a similar  fine 
structure. The calculations of the energy spectrum and tunneling in a tilted 
magnetic field are essential to understanding of peculiarities observed.  To 
the authors' knowledge, such calculations for structures based on Kane 
semiconductors are not available.

\section{Conclusion}
\label{five}
	Tunneling conductivity oscillations in a magnetic field have been 
investigated for structures based on narrow-gap $\text{Hg}_{1-
x}\text{Cd}_x\text{Te}$.  
These oscillations in the structure investigated are shown to arise from 
tunneling into bulk Landau levels.  Rather 
remarkably,  at $B\perp n$ the relationship of the amplitudes of the maxima 
associated with tunneling into different spin states varies
significantly for the  structures investigated, so that only
tunneling into one-spin sublevels is  
observed  in some structures. 

	To understand the causer of this distinction, we have calculated 
the behavior of the Landau levels near the surface and the tunneling 
conductivity  oscillation at $B\perp n$. The relationship of the amplitudes of 
the maxima associated with tunneling into different spin states is shown to 
depend dramatically on the surface potential. Thus the investigations of the 
tunneling 
conductivity at $B\perp n$ provide a way to estimate  the value of the 
surface potential when this is small and cannot localize the 2D states.

\begin{figure}
\caption{Energy diagram of metal -- insulator -- semiconductor structure. $E_c$ and $E_v$ are the energies of the edges of conduction and valence bands, respectively.} 
\label{fig1}
\end{figure}

\begin{figure}
\caption{Oscillations of tunneling conductivity in a magnetic field at 
$V=130$ mV for various magnetic field orientations for structures I (a) and 
II (b).}
\label{fig2}
\end{figure}

\begin{figure}
\caption{The bias dependences of tunneling conductivity in  magnetic field 
4 T for structures I (a) and II (b) for two orientations of magnetic field.}
\label{fig3}
\end{figure}

\begin{figure}
\caption{The model of the energy structure of insulator -- semiconductor 
boundary used for calculations.}
\label{fig4}
\end{figure}

\begin{figure}
\caption{The $x_0$ dependences of the energy of the Landau sublevels in 
vicinity of the boundary. The parameters used are given in the text, $B=4$ 
T.}
\label{fig5}
\end{figure}

\begin{figure}
\caption{The $x_0$ dependences of $\left(\hat{v}\Psi\right)^2$ for 
$\text{\em a}_1$ and  $\text{\em b}_1$ Landau sublevels calculated with 
the same parameters as in Fig. \protect\ref{fig5}, $B=4$ T.}
\label{fig6}
\end{figure}

\begin{figure}
\caption{The contributions to the tunneling conductivity of various Landau 
levels, $B=4$ T.}
\label{fig7}
\end{figure}

\begin{figure}
\caption{The oscillatory part of the tunneling conductivity calculated for 
two orientations of magnetic field, $B=4$ T.}
\label{fig8}
\end{figure}

\begin{figure}
\caption{The $x_0$ dependence of the energy of $\text{\em a}_1$ and  
$\text{\em b}_1$ Landau sublevels calculated for three values of the 
surface potential.}
\label{fig9}
\end{figure}

\begin{figure}
\caption{Oscillatory part of the tunneling conductivity for three values of 
the surface potential.}
\label{fig10}
\end{figure}

\begin{figure}
\caption{The fine structure of the tunneling conductivity oscillations in 
tilted magnetic field for tunnel structure based on $\text{p-Hg}_{1-
x}\text{Cd}_x\text{Te}$ with $E_g=20$ meV. $B=5.5$ T}
\label{fig11}
\end{figure}

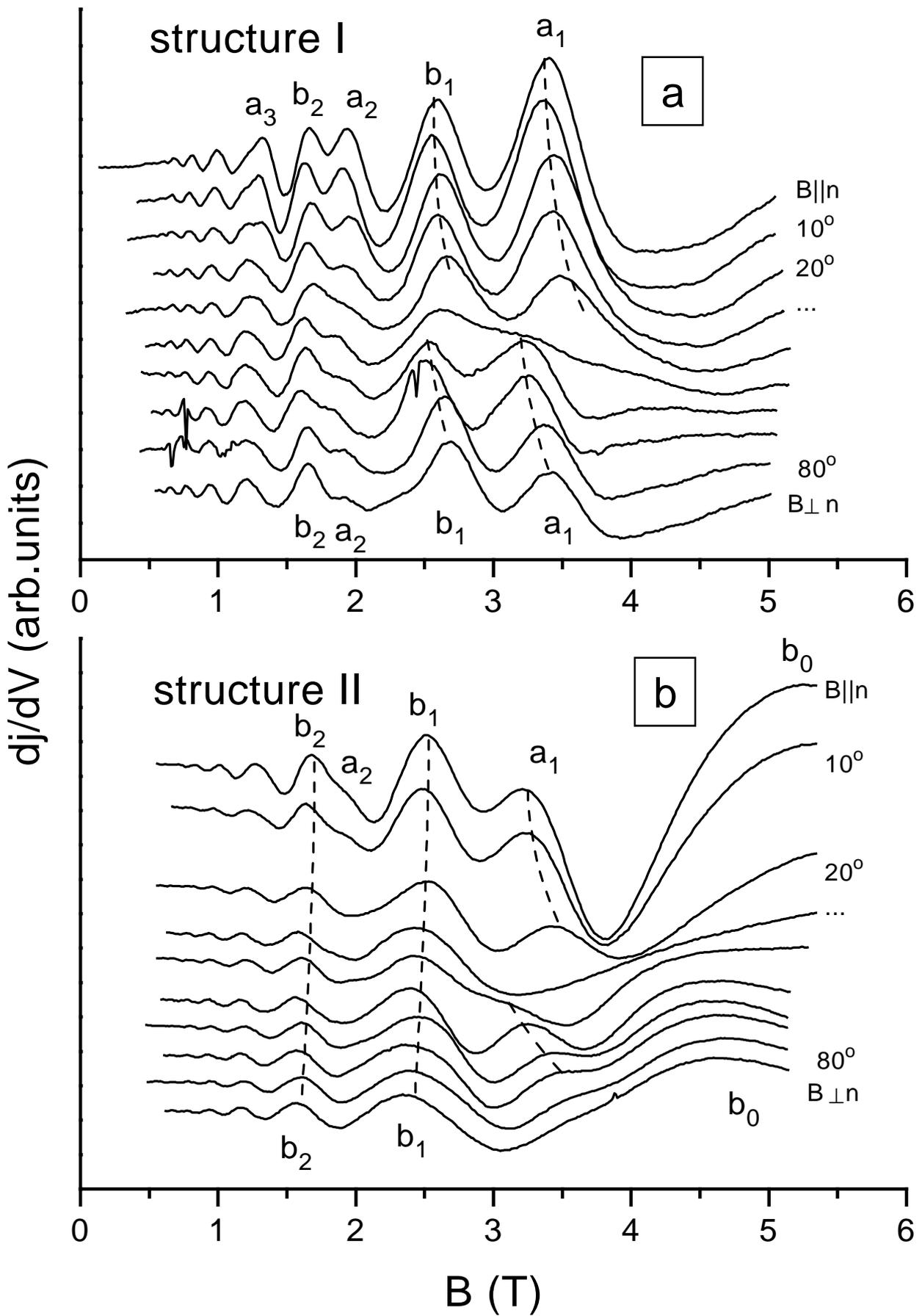
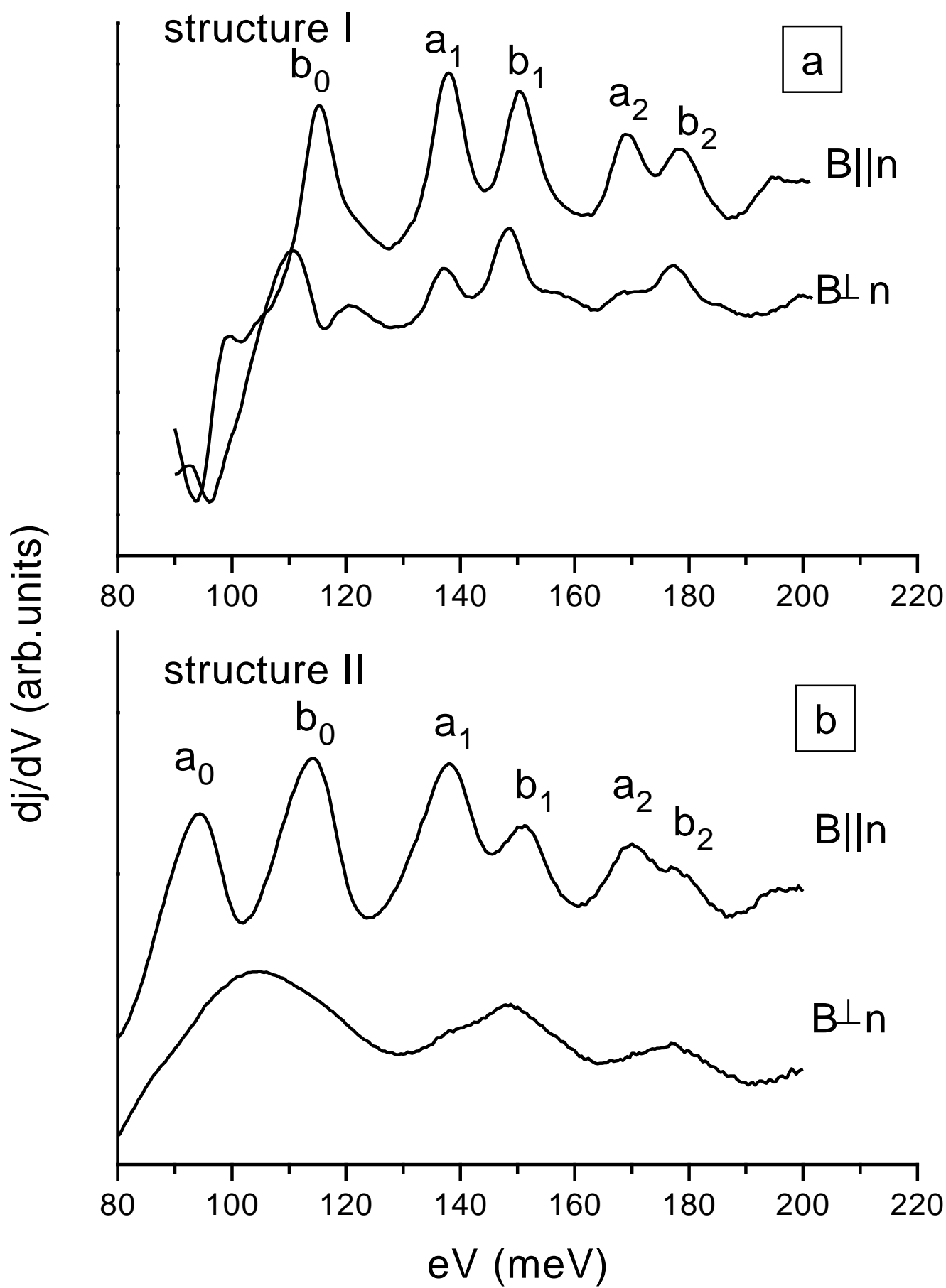
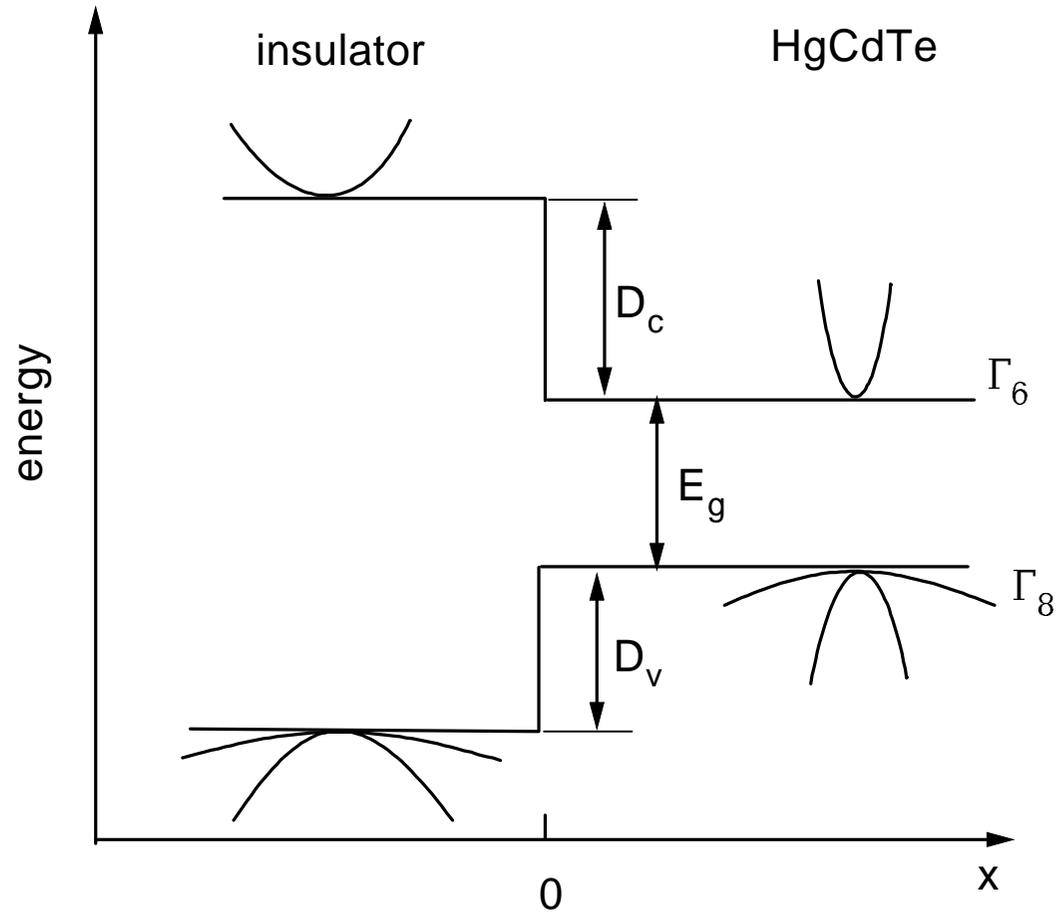
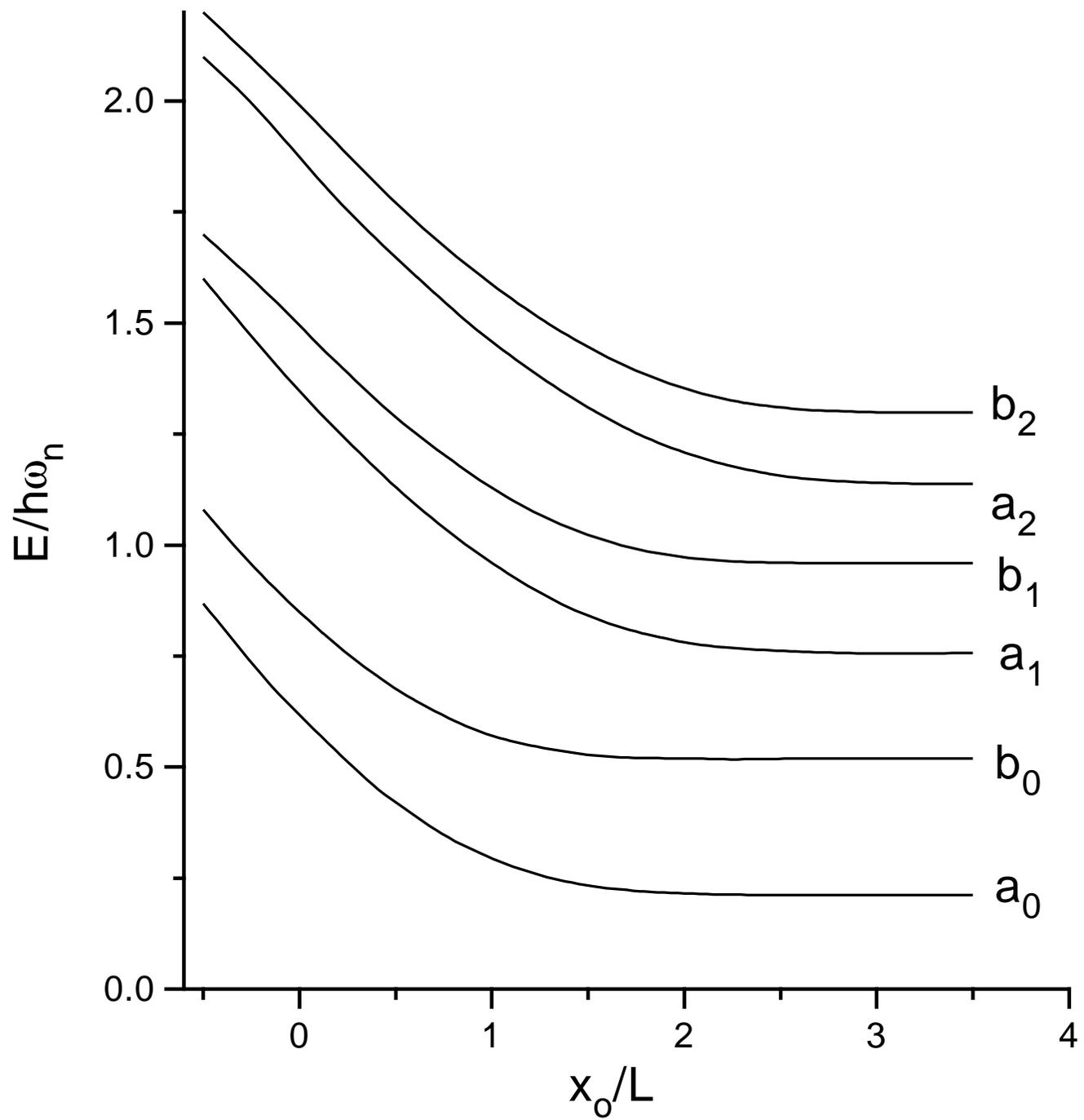
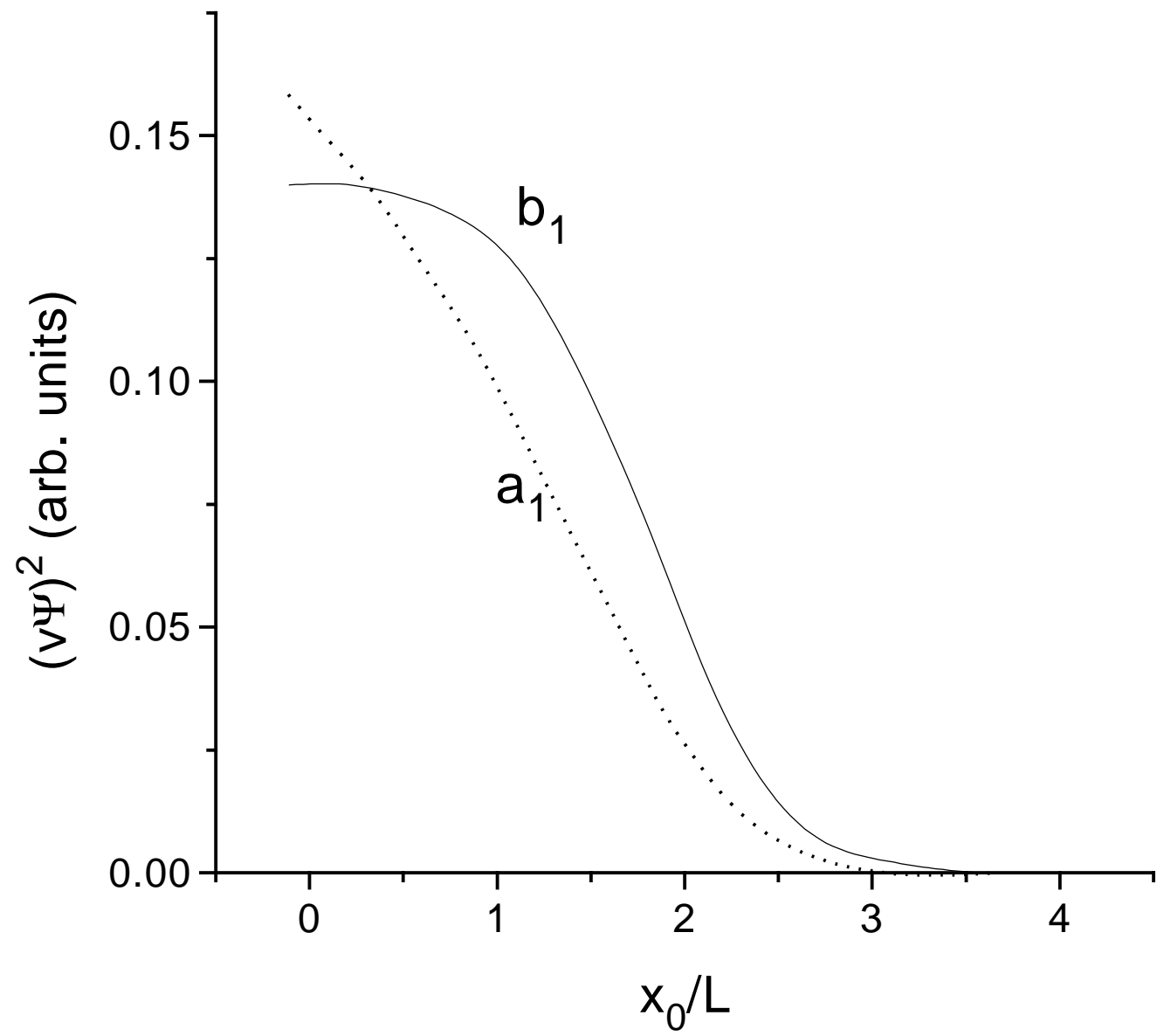
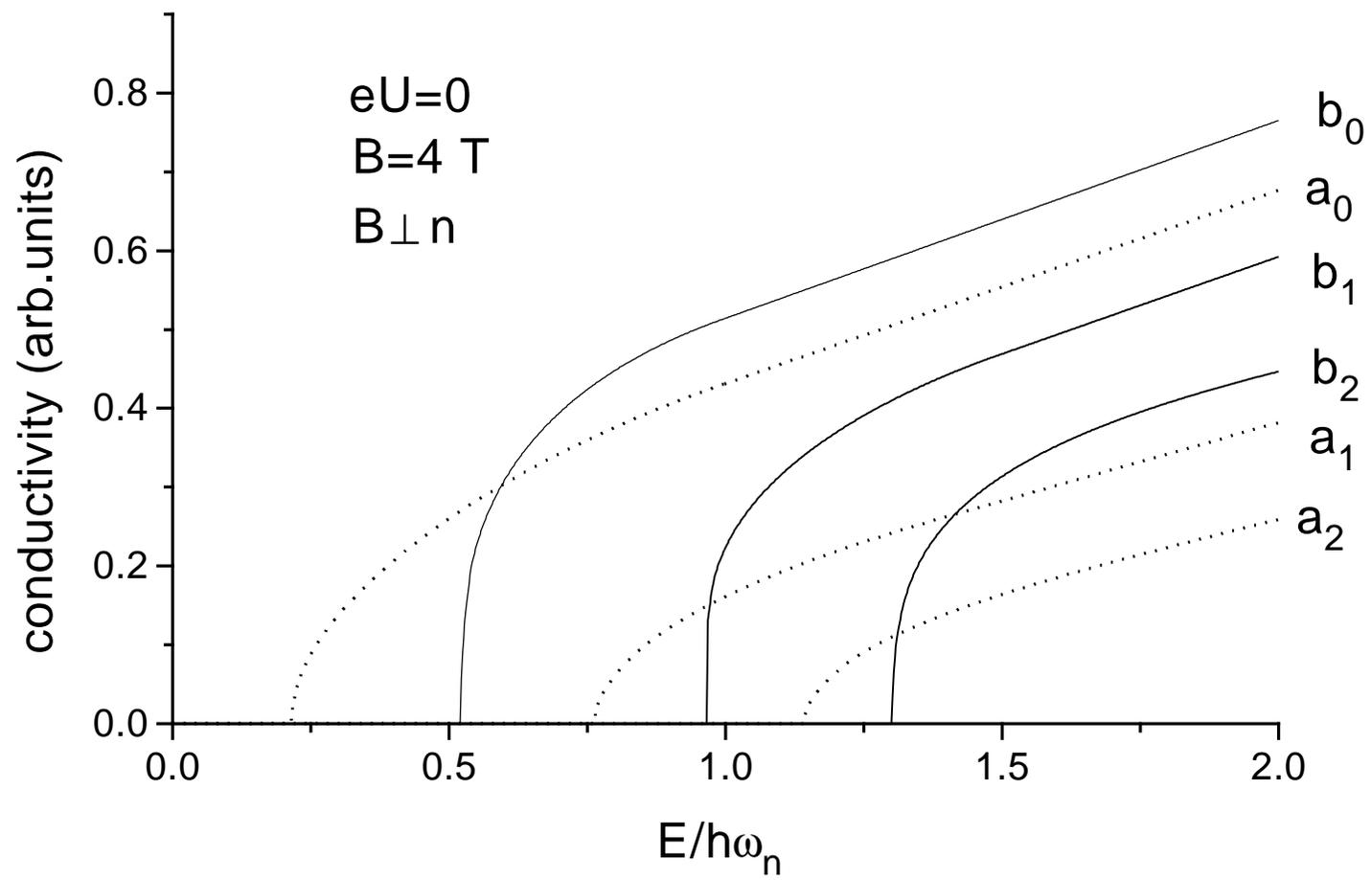
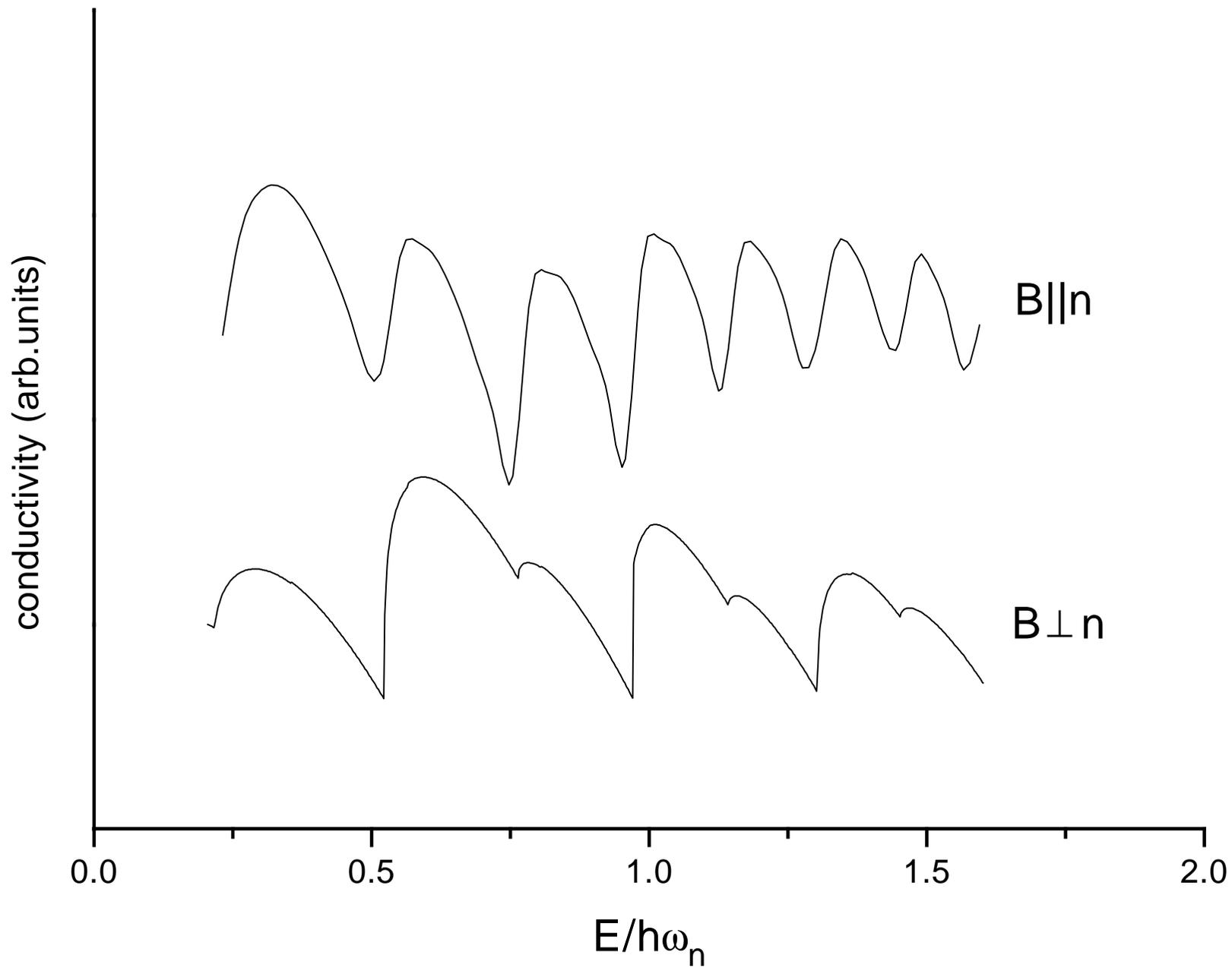
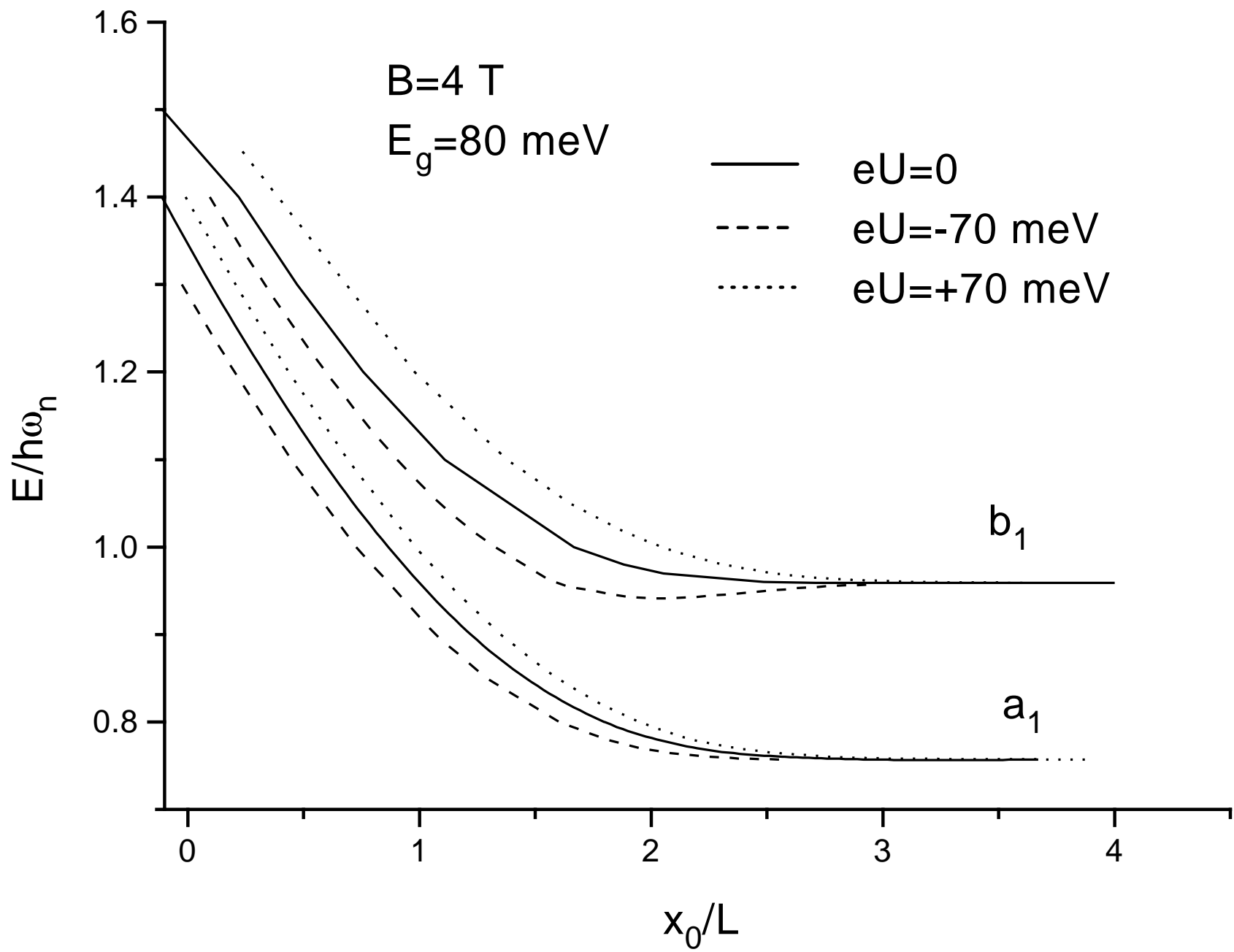
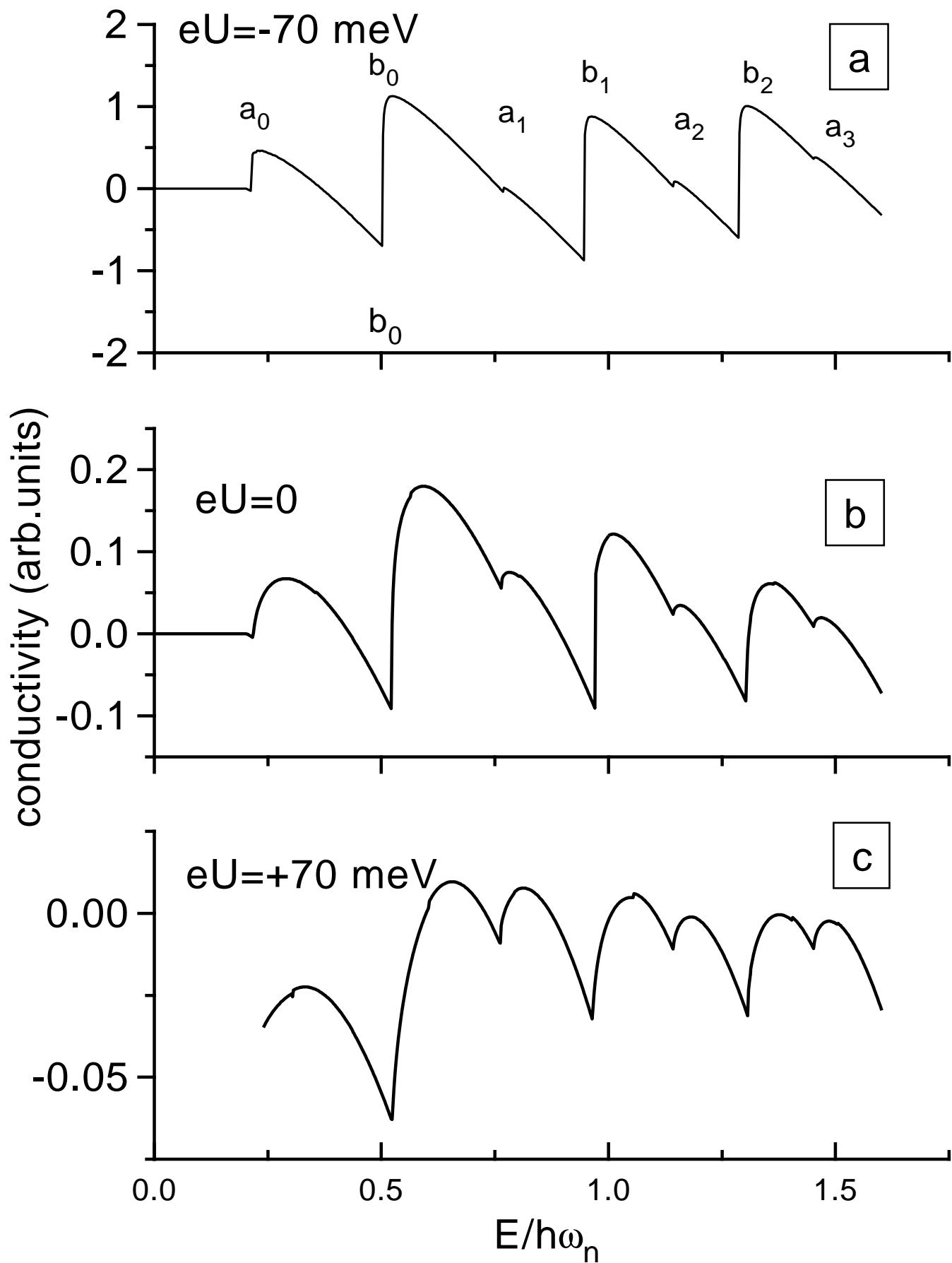
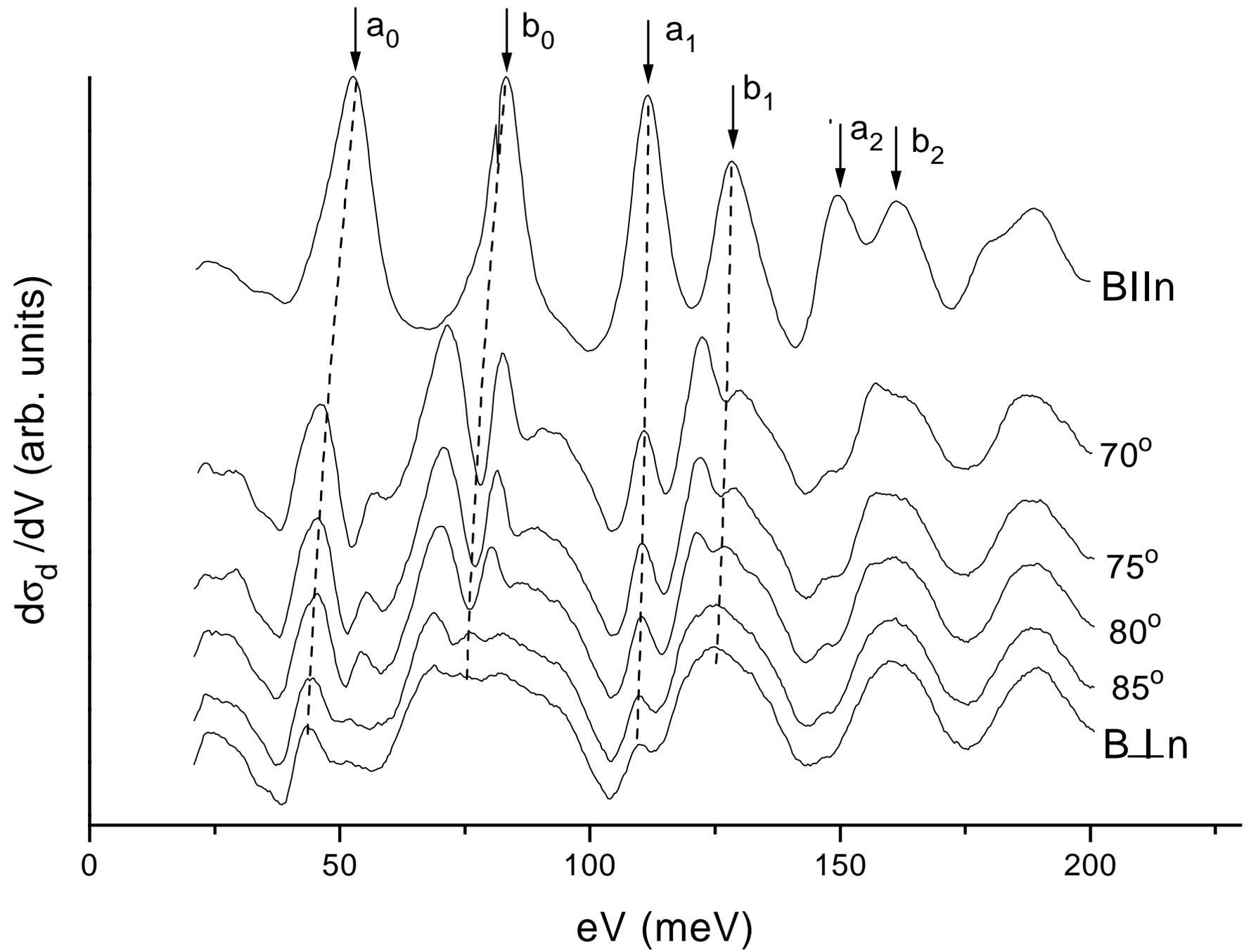
\end{document}